\documentclass[paper]{JHEP3} 

\usepackage{amsfonts}
\usepackage{amsmath}
\usepackage{epsfig}
\usepackage{latexsym}
\usepackage{graphicx}
\usepackage{subfigure}
\usepackage{amssymb}
\numberwithin{equation}{section}
\allowdisplaybreaks

\def\be{\begin{equation}}
\def\ee{\end{equation}}
\def\bea{\begin{eqnarray}}
\def\eea{\end{eqnarray}}
\def\bequ{\begin{equation}}
\def\eequ{\end{equation}}

\newcommand{\eq} {equation}
\newcommand{\eqa} {eqnarray}
\newcommand{\NN} {\mbox {$\nonumber$}}

\title{
\Large Direct derivation of ``mirror'' ABJ partition function}

\author{
{\large Masazumi Honda }\vspace*{0.5cm} \\
High Energy Accelerator Research Organization (KEK),\\
Tsukuba, Ibaraki 305-0801, Japan, \\
Yukawa Institute for Theoretical Physics, Kyoto University,\\
Kitashirakawa Oiwakecho, Sakyo-ku, Kyoto 606-8502, Japan

\vspace*{0.5cm} \\
\email{mhonda@post.kek.jp}}

\preprint{KEK-TH-1671,YITP-13-95}

\abstract{
We study 
the partition function of the three-dimensional $\mathcal{N}=6\ U(N)_k \times U(N+M)_{-k}$ superconformal Chern-Simons matter theory 
known as the ABJ theory.
We prove that
the ABJ partition function on $S^3$ is exactly the same as a formula
recently proposed by Awata, Hirano and Shigemori.
While this formula was previously obtained by an analytic continuation from the $L(2,1)$ lens space matrix model,
we directly derive this by using a generalization of the Cauchy determinant identity.
We also give an interpretation for the formula from brane picture.
}


\begin{document}
\setcounter{footnote}{0}

\section{Introduction}
\label{sec:intro}
Recently there has been significant progress in understanding low-energy effective theories of multiple M2-branes.
The simplest case of such theories is the so-called ABJM theory \cite{Aharony:2008ug},
which is the 3d $\mathcal{N}=6$ supersymmetric Chern-Simons matter theory (CSM) with the gauge group $U(N)_k \times U(N)_{-k}$.
The authors in \cite{Aharony:2008ug} discussed that
this theory describes $N$ M2-branes on $\mathbb{C}^4 /\mathbb{Z}_k$ in the low-energy limit.
Furthermore it has been shown 
by the localization method \cite{Pestun:2007rz,Kapustin:2009kz} (see also \cite{Nian:2013qwa})
that a class of supersymmetric observables in $\mathcal{N}=2$ theory on $S^3$ 
have representations in terms of certain matrix integrals.
Thanks to the localization technique, 
several works have extensively studied
the partition function and BPS Wilson loops in the ABJM theory on $S^3$ \cite{Marino:2009jd,Drukker:2009hy,Herzog:2010hf,Drukker:2011zy,Okuyama:2011su,Marino:2011eh,Hanada:2012si,Klemm:2012ii,Hatsuda:2012hm,Bhattacharyya:2012ye,Hatsuda:2012dt,Grassi:2013qva,Hatsuda:2013yua,Marino:2011nm}.
Especially a breakthrough was caused by a seminal paper \cite{Marino:2011eh}, 
which rewrites the ABJM partition function 
as an ideal Fermi gas system (see also \cite{Kapustin:2010xq,Herzog:2010hf,Okuyama:2011su,Marino:2012az}).
Based on this formalism, 
recent studies have revealed 
structures of the partition function \cite{Hatsuda:2012dt} and half-BPS Wilson loop \cite{Hatsuda:2013yua} 
including worldsheet and membrane (D2-brane) instanton effects \cite{Cagnazzo:2009zh,Drukker:2011zy}.

In this paper we study the $\mathcal{N}=6$ CSM with more general gauge group $U(N)_k \times U(N+M)_{-k}$
known as the ABJ theory \cite{Aharony:2008gk}.
This theory has been expected to arise
when we have $N$ M2-branes on $\mathbb{C}^4 /\mathbb{Z}_k$,
together with $M$ fractional M2-branes sitting at the singularity.
The authors in \cite{Aharony:2008gk} also argued that
the ABJ theory has good approximations by the 11d SUGRA on $AdS_4 \times S^7 /\mathbb{Z}_k$
with discrete torsion for $N^{1/5}\gg k$, and
the type IIA SUGRA on $AdS_4 \times \mathbb{CP}^3$ with nontrivial B-field holonomy for $N^{1/5}\ll k \ll N$, respectively.
Furthermore the recent works \cite{Chang:2012kt} have also conjectured that
the ABJ theory is dual to $\mathcal{N}=6$ parity-violating Vasiliev theory on $AdS_4$ with a $U(N)$ gauge symmetry
when $M,k\gg 1$ with $M/k$ and $N$ kept fixed.
Thus it is worth studying the ABJ theory in detail.

Here we study 
the partition function of the $U(N)_k \times U(N+M)_{-k}$ ABJ theory on $S^3$.
By using the localization method,
the partition function is given by \cite{Kapustin:2009kz}
\begin{\eqa}
Z_{\rm ABJ}^{(N,N+M)}(k) 
&=& \frac{i^{-\frac{1}{2}(N^2 -(N+M)^2 ){\rm sign}(k)}}{(N+M)!N!} \int_{-\infty}^\infty \frac{d^{N+M} \mu}{(2\pi )^{N+M}} \frac{d^N \nu}{(2\pi )^N}
e^{-\frac{ik}{4\pi}\left( \sum_{j=1}^{N+M} \mu_j^2 -\sum_{a=1}^N \nu_a^2 \right) } \NN\\
&&\times \Biggl[ 
  \frac{ \prod_{1\leq j<l \leq N+M}2\sinh{\frac{\mu_j -\mu_l }{2}} \prod_{1\leq a<b \leq N}2\sinh{\frac{\nu_a -\nu_b }{2}}}
        {\prod_{j=1}^{N+M} \prod_{b=1}^N 2\cosh{\frac{\mu_j -\nu_b }{2}}} \Biggr]^2 .
\label{starting}
\end{\eqa}
For $M\neq  0$, this representation is not suitable for the Fermi gas approach
since the Cauchy identity is not helpful in contrast to the ABJM case \cite{Marino:2011eh}.
Nevertheless, Awata, Hirano and Shigemori (AHS) recently proposed \cite{Awata:2012jb}
that the ABJ partition function is equivalent to 
\begin{\eqa}
Z_{\rm AHS}^{(N,N+M)}(k)
&=& \frac{i^{-\frac{1}{2}(N^2 +(N+M)^2 ){\rm sign}(k) +N+\frac{M}{2}}  
             (-1)^{\frac{N}{2}(N-1)}}{2^N k^{N+M/2}N!} (1-q)^{\frac{M(M-1)}{2}} G_2 (M+1 ;q) \NN\\
&& \int_{-i \infty -2\pi\eta}^{i \infty -2\pi\eta} \frac{d^N s}{(-2\pi i )^N} 
\prod_{a=1}^N \frac{1}{2\sin{\frac{s_a}{2}}} \frac{(q^{\frac{s_a}{2\pi} +1})_M}{(-q^{\frac{s_a}{2\pi} +1})_M} 
\prod_{1\leq a<b \leq N}\frac{(1 -q^{\frac{s_b -s_a }{2\pi}} )^2}{(1 +q^{\frac{s_b -s_a }{2\pi}} )^2} ,
\label{AHS}
\end{\eqa}
where 
\[
q=e^{-\frac{2\pi i}{k}},\ 
(a)_n = \prod_{m=0}^{n-1} (1 -aq^m ),\ 
G_2 (z+1;q ) 
= (1-q)^{-\frac{z}{2}(z-1)}  \prod_{m=1}^\infty \Biggl[ \left(  \frac{1-q^{z+m}}{1-q^m}  \right)^m (1-q^m )^z \Biggr] .
\]
Here $\eta $ specifies the integral contour.
The authors in \cite{Awata:2012jb} have determined $\eta$ for $N=1$ as
\begin{\eq}
\eta =\left\{ \begin{matrix}
0_+ & {\rm for} & \frac{|k|}{2}-M\geq 0 \cr
-\frac{|k|}{2}+M+ 0_+ & {\rm for} & \frac{|k|}{2}-M\leq 0 \cr
\end{matrix}\right.  ,
\label{eta_choice}
\end{\eq}
to be consistent with the Seiberg-like duality \cite{Aharony:2008gk}
but not for general $N$.

One expects that
the AHS formula \eqref{AHS} gives a generalization of the ``mirror'' description of the ABJM partition function. 
One of the strongest evidence is that
$Z_{\rm AHS} |_{M=0,k=1}$ 
is the same as the partition function of the $\mathcal{N}=4$ super QCD 
with one adjoint and fundamental hypermultiplets \cite{Kapustin:2010xq} 
related through 3d mirror symmetry \cite{Intriligator:1996ex,Hanany:1996ie,Aharony:2008ug}.
There is also an interpretation for general $k$ 
from the S-dual brane construction \cite{Gulotta:2011si,Benvenuti:2011ga}.

The AHS representation also has several advantages.
First, this is suitable for the Fermi gas approach and Tracy-Widom theorem \cite{Tracy:1995ax}, 
which reduces the grand canonical analysis to Thermodynamic Bethe Ansatz-like equation.
Second, it is easier to perform Monte Carlo simulation as in the ABJM case \cite{Hanada:2012si} than the original formula \eqref{starting}.
Finally, the AHS formula highly simplifies analysis in the Vasiliev limit.
Despite of the advantages, the AHS proposal is still conjecture in the following senses:
\begin{itemize}
\item The derivation of $Z_{\rm AHS}$ started with
an analytic continuation \cite{Marino:2009jd} from the partition function of the $L(2,1)$ lens space matrix model \cite{Marino:2002fk}.
The analytic continuation has not been rigorously justified 
in spite of much strong evidence \cite{Marino:2009jd,Drukker:2011zy,Hanada:2012si,Hatsuda:2012dt,Yost:1991ht}.

\item While we can represent the partition function of the $L(2,1)$ matrix model in terms of a convergent series,
its analytic continuation to the ABJ theory yields a non-convergent series.
The AHS formula corresponds to its well-defined integral representation
and reproduces the series order by order in the perturbative expansion by $2\pi i/k$
although there would be non-perturbative ambiguity generically.
\end{itemize}

In this paper we prove the AHS conjecture $Z_{\rm ABJ}=Z_{\rm AHS}$ and
determine the integral contour for arbitrary parameters as discussed in section~\ref{sec:proof}.
It will turn out that the choice \eqref{eta_choice} of $\eta$  
is still correct even for general $N$.
Section~\ref{sec:con} is devoted to discussion.

\section{Proof}
\label{sec:proof}
In this section we prove $Z_{\rm ABJ}=Z_{\rm AHS}$.
Let us start with the localization formula \eqref{starting}.
For $M=0$, the Cauchy determinant identity is quite useful 
to derive its ``mirror'' description \cite{Kapustin:2010xq,Okuyama:2011su,Marino:2011eh}, but not for $M\neq 0$.
Here instead we use a generalization\footnote{
We thank Sanefumi Moriyama to tell us about this identity 
and suggest that the identity would be useful for analyzing the ABJ matrix model \eqref{starting}.
He showed us a different proof for the identity in April 2012.
In the meanwhile, we remembered the identity 
when we read a (Japanese) textbook on conformal field theory written by Yasuhiko Yamada.
Therefore we are also grateful to Yasuhiko Yamada.
} of the Cauchy identity (corresponding to Lemma.~2 of \cite{Forrester}) :
\begin{\eqa}
&&\frac{\prod_{j<l}(x_j -x_l ) \prod_{a<b}(y_a -y_b )}{\prod_{j,b}(x_j -y_b )} \NN\\
&=&(-1)^{-\frac{N}{2}(N-1)}  \left| \begin{matrix}
    x_1^{M-1} & x_1^{M-2} & \cdots & 1          & \frac{1}{x_1 -y_1} & \frac{1}{x_1 -y_2}  & \cdots & \frac{1}{x_1 -y_N} \cr
    x_2^{M-1} & x_2^{M-2} & \cdots & 1          & \frac{1}{x_2 -y_1} & \frac{1}{x_2 -y_2}  & \cdots & \frac{1}{x_2 -y_N} \cr
      \vdots   &   \vdots   &  \vdots & \vdots  & \vdots              &  \vdots                &\vdots  & \vdots               \cr
    x_{N+M}^{M-1} & x_{N+M}^{M-2} & \cdots & 1     & \frac{1}{x_{N+M} -y_1} & \frac{1}{x_{N+M} -y_2}  & \cdots & \frac{1}{x_{N+M} -y_N} \cr
\end{matrix}\right| ,
\label{det}
\end{\eqa}
where $j,l,a$ and $b$ run $1\leq j,l \leq N+M$, $1\leq a,b \leq N$.  
One of easiest way to prove this identity is to use Boson-Fermion correspondence in 2d CFT.
More concretely, the left-hand side corresponds to a correlation function on $\mathbb{P}^1$ of a bc system in ``Boson'' representation:
\begin{\eq}
\langle M | b (x_1 ) \cdots b(x_{N+M}) c(y_N ) \cdots c(y_1 ) | 0\rangle ,
\end{\eq}
where we have identified as $b \leftrightarrow :e^{\varphi} :$ and $c \leftrightarrow :e^{-\varphi} :$ by using a free boson satisfying
\begin{\eqa}
&&\varphi (z) = \tilde{q} +a_0 \log{z} -\sum_{n\neq 0} \frac{a_n}{n} z^{-n} ,\quad
[a_m ,\tilde{q}]=\delta_{m,0} ,\quad [a_m ,a_n ] =m\delta_{m+n,0} ,\NN\\
&& a_{n\geq 0} |0\rangle =0 ,\quad  \langle 0| \tilde{q} =\langle 0| a_{n<0} =0 ,\quad \langle 0|0\rangle =1,\quad 
|M\rangle  = e^{M\tilde{q}} |0\rangle ,
\end{\eqa}
while the right-hand side is the one in ``Fermion'' representation with identifications: $b \leftrightarrow \bar{\psi}$ and $c \leftrightarrow \psi$
in terms of charged fermions satisfying
\begin{\eqa}
&&\bar{\psi} (z) = \sum_{n\in \mathbb{Z}+1/2 } \bar{\psi}_n z^{-n-\frac{1}{2}} ,\quad
\psi (z) = \sum_{n\in \mathbb{Z}+1/2 } \psi_n z^{-n-\frac{1}{2}} ,\NN\\
&&\{ \bar{\psi}_m ,\bar{\psi}_n \} =\{ \psi_m ,\psi_n \} = 0 ,\quad \{ \bar{\psi}_m ,\psi_n \} =\delta_{m+n,0} ,\NN\\
&& \bar{\psi}_{n> 0} |0\rangle =\psi_{n> 0} |0\rangle =0 ,\quad 
 \langle 0| \bar{\psi}_{n<0} =\langle 0| \psi_{n<0} =0 ,\quad
|M\rangle  = \bar{\psi}_{-M+\frac{1}{2}} \cdots \bar{\psi}_{-\frac{1}{2}} |0\rangle .
\end{\eqa}
If we take $x_j =e^{\mu_j}$ and $y_a =-e^{\nu_a}$ in the determinant identity \eqref{det}, then we find
\begin{\eqa}
&&\frac{ \prod_{j<l}2\sinh{\frac{\mu_j -\mu_l }{2}} \prod_{a<b}2\sinh{\frac{\nu_a -\nu_b }{2}}}{\prod_{j,b}2\cosh{\frac{\mu_l -\nu_b }{2}}} \NN\\
&&= \prod_{j=1}^{N+M} e^{-M \frac{\mu_j}{2}} \prod_{a=1}^N e^{M \frac{\nu_a}{2}} 
   \det{\left(  \frac{\theta_{N,l}}{2\cosh{\frac{\mu_j -\nu_l }{2}}}  +e^{\left( N+M+\frac{1}{2} -j\right) \mu_j}  \theta_{l,N+1} \right)}  ,
\label{determinant}
\end{\eqa}
where
\begin{\eq}
\theta_{j,l}=
\left\{ \begin{matrix}
 1 & {\rm for} & j\geq l  \cr
 0 & {\rm for} & j< l  \cr
\end{matrix} \right.   .
\end{\eq}
Plugging this into \eqref{starting}, we find
\begin{\eqa}
Z_{\rm ABJ}^{(N,N+M)}(k) 
&=& \frac{\mathcal{N}_{\rm ABJ}}{N!} \sum_\sigma (-1)^\sigma \int_{-\infty}^\infty \frac{d^{N+M} \mu}{(2\pi )^{N+M}} \frac{d^N \nu}{(2\pi )^N}
\prod_{j=1}^{N+M} e^{-\frac{ik}{4\pi}\mu_j^2-M \mu_j}
\prod_{a=1}^N \frac{e^{\frac{ik}{4\pi}\nu_a^2 +M\nu_a}}{2\cosh{\frac{\mu_a -\nu_a }{2}}}  \NN\\
&& \prod_{l=N+1}^{N+M} e^{\left( N+M+\frac{1}{2} -l\right) \mu_l}
\prod_{j=1}^{N+M} \left(  \frac{\theta_{N,j}}{2\cosh{\frac{\mu_{\sigma (j)} -\nu_j }{2}}}  
                                 +e^{\left( N+M+\frac{1}{2} -j\right) \mu_{\sigma (j)}}\theta_{j,N+1} \right)  , \NN
\end{\eqa}
where
\begin{\eq}
\mathcal{N}_{\rm ABJ} = i^{-\frac{1}{2}(N^2 -(N+M)^2 ){\rm sign}(k)} .
\end{\eq}
Making a Fourier transformation
\begin{\eq}
\frac{1}{2\cosh{\frac{p}{2}}} = \frac{1}{\pi} \int_{-\infty}^\infty dx \frac{e^{\frac{i}{\pi}px}}{2\cosh{x}},
\end{\eq}
and introducing auxiliary variables $y_{N+1} ,\cdots ,y_{N+M}$ constrained by
\begin{\eq}
y_l =\frac{\pi}{i}\left( N+M+\frac{1}{2} -l \right) \quad
{\rm with}\ l=N+1,\cdots ,N+M ,
\end{\eq} 
the partition function becomes 
\begin{\eqa}
Z_{\rm ABJ}
&=& \frac{\mathcal{N}_{\rm ABJ}}{N!} \sum_\sigma (-1)^\sigma 
\int_{-\infty}^\infty \frac{d^N x}{\pi^N} \frac{d^{N+M} y}{\pi^{N+M}} \frac{d^{N+M} \mu}{(2\pi )^{N+M}} \frac{d^N \nu}{(2\pi )^N}
\prod_{j=1}^{N+M} e^{-\frac{ik}{4\pi}\mu_j^2-M \mu_j} \NN\\ 
&&\times \prod_{a=1}^N \frac{e^{\frac{ik}{4\pi}\nu_a^2 +M\nu_a +\frac{i}{\pi}x_a (\mu_a -\nu_a )+\frac{i}{\pi}( y_{\sigma (a)} \mu_a -y_a \nu_a )}}
                                 {2\cosh{x_a} \cdot 2\cosh{y_a}} \NN\\
&&\times \prod_{l=N+1}^{N+M} \Biggl[ 
   \pi \delta \left( y_l -\frac{\pi}{i}(N+M+1/2 -l)  \right) e^{\left( N+M+\frac{1}{2} -l\right) \mu_l +\frac{i}{\pi} y_{\sigma (l)} \mu_l } \Biggr] .
\end{\eqa}
Here we used 
$\sum_{j=1}^{N+M} y_j \mu_{\sigma (j)} =\sum_{j=1}^{N+M} y_{\sigma^{-1} (j)} \mu_j$
and redefined the permutation symbol as $\sigma^{-1} \rightarrow \sigma$.
Performing the Fresnel integrals over $\mu_i$ and $\nu_a$ allows us to find
\begin{\eqa}
Z_{\rm ABJ} 
&=& \frac{e^{-\frac{iM \pi}{4} {\rm sign}(k) } \mathcal{N}_{\rm ABJ}}{N! |k|^{N+\frac{M}{2}}} 
  \sum_\sigma (-1)^\sigma  \int_{-\infty}^\infty \frac{d^N x}{\pi^N} \frac{d^{N+M} y}{\pi^{N+M}}   
 \prod_{a=1}^N \frac{e^{ -\frac{2i}{k\pi}x_a (y_a -y_{\sigma (a)}) +\frac{2}{k}M(y_a -y_{\sigma (a)})} }  {2\cosh{x_a} \cdot 2\cosh{y_a}} \NN\\
&& \prod_{l=N+1}^{N+M} \Biggl[ \pi   e^{  -\frac{i\pi}{k}\left( N+\frac{1}{2} -l  \right)^2}  \delta \left( y_l -\frac{\pi}{i}(N+M+1/2 -l)  \right)
                        e^{\frac{i}{k\pi}y_l^2  +\frac{2}{k} \left( N+\frac{1}{2} -l \right) y_{\sigma (l)} } \Biggr] . 
\label{turning}
\end{\eqa}
Note that the integration over $x_a$ is convergent only for\footnote{
The saturated case $\left| \frac{2}{k\pi} {\rm Im}(y_a -y_{\sigma (a)} )\right| = 1$ is understood as a limit 
from $\left| \frac{2}{k\pi} {\rm Im}(y_a -y_{\sigma (a)} )\right| < 1$.
} $\left| \frac{2}{k\pi} {\rm Im}(y_a -y_{\sigma (a)} )\right| \leq 1$.
Since $y_{\sigma (a)}$ would be $-i\pi (M-1/2)$ depending on the permutation,
the integration is always safe for $2M \leq |k| +1$.
On the other hand, a part of the integrations is divergent for $2M > |k| +1$.
However, these divergences must be apparent and cancel out after summing over the permutation
since this case is equivalent to the safe case through the Seiberg-like duality \cite{Aharony:2008gk}
between the ABJ theories with gauge groups 
\begin{\eq}
U(N)_k \times U(N+M)_{-k}\quad {\rm and} \quad U(N+|k|-M)_k \times U(N)_{-k},
\label{Seiberg}
\end{\eq}
which has been proven for the $S^3$ partition functions \cite{Willett:2011gp,Kapustin:2010mh}.
Although we could regularize the divergences,
instead we will adopt another way as discussed in sec.~\ref{sec:latter}.

\subsection{For $2M\leq |k|+1$}
\label{sec:former}
We can continue our computation straightforwardly for this case. 
Integrating over $x_a$ leads us to
\begin{\eqa}
Z_{\rm ABJ}^{(N,N+M)}(k) 
&&= \frac{ e^{-\frac{iM \pi}{4} {\rm sign}(k) } \mathcal{N}_{\rm ABJ}}{N! |k|^{N+\frac{M}{2}}} 
  \sum_\sigma (-1)^\sigma  \int_{-\infty}^\infty \frac{d^{N+M} y}{\pi^{N+M}}   
 \prod_{a=1}^N \frac{e^{ \frac{2}{k}M(y_a -y_{\sigma (a)})}}  {2\cosh{\frac{y_a -y_{\sigma (a)}}{k}} \cdot 2\cosh{y_a}} \NN\\
&& \prod_{l=N+1}^{N+M} \Biggl[ \pi   e^{  -\frac{i\pi}{k}\left( N+\frac{1}{2} -l  \right)^2}  \delta \left( y_l -\frac{\pi}{i}(N+M+1/2 -l)  \right)
                        e^{\frac{i}{k\pi}y_l^2  +\frac{2}{k} \left( N+\frac{1}{2} -l \right) y_{\sigma (l)} } \Biggr]  .\NN
\end{\eqa}
Noting $\sum_a (y_a -y_{\sigma (a)}) = -\sum_{l=N+1}^{N+M} (y_l -y_{\sigma (l)} )$ and rescaling $y_a$ as $y_a \rightarrow y_a /2$,
one finds
\begin{\eqa}
Z_{\rm ABJ} 
&=&  \frac{e^{-\frac{iM \pi}{4} {\rm sign}(k) } \mathcal{N}_{\rm ABJ} }{N! |k|^{N+\frac{M}{2}}}  \int_{-\infty}^\infty \frac{d^{N+M} y}{(2\pi )^{N+M}}   
 \prod_{a=1}^N \frac{1}{2\cosh{\frac{y_a}{2}}} \NN\\
&&\times \prod_{l=N+1}^{N+M} \Biggl[ \pi   e^{  -\frac{i\pi}{k}\left( N+\frac{1}{2} -l  \right)^2}  \delta \left( \frac{y_l}{2} -\frac{\pi}{i}(N+M+1/2 -l)  \right)
                        e^{\frac{i}{4k\pi}y_l^2  -\frac{M}{k}y_l } \Biggr] \NN\\
&&\times    \det{\left(  \frac{\theta_{N,l}}{2\cosh{\frac{y_j -y_l}{2k}}}  +e^{\frac{1}{k}(N+M+1/2 -l)y_j}  \theta_{l,N+1} \right)} .
\end{\eqa}
If we use the identity \eqref{determinant} again and integrate over $y_{N+1},\cdots ,y_{N+M}$, 
then we obtain
\begin{\eqa}
&&Z_{\rm ABJ}^{(N,N+M)}(k) \NN\\
&&=    \frac{i^{-\frac{{\rm sign}(k)}{2}(N^2 +(N+M)^2 ) } (-1)^{\frac{N}{2}(N-1)+\frac{M}{2}(M-1)+NM} i^{N+\frac{M}{2}}   }{ N! 2^N k^{N+\frac{M}{2}}} 
q^{\frac{M}{12}(M^2 -1)} \prod_{1 \leq l<m \leq M}\Bigl[ 2i \sin{\frac{\pi (l-m)}{k}} \Bigr]   \NN\\
&&\int_{-\infty}^\infty \frac{d^N y}{(2\pi )^N}   
    \prod_{a<b} \tanh^2{\frac{y_a -y_b }{2k}} 
\prod_{a=1}^N \Biggl[ \frac{1}{2\cosh{\frac{y_a}{2}}} \prod_{l=0}^{M-1}  \tanh{\frac{y_a +2\pi i (l +1/2 )  }{2k}}  \Biggr] .
\label{ABJ_mirror}
\end{\eqa}
Taking account of
\begin{\eqa}
&&\prod_{1 \leq l<m \leq M}\Bigl[ 2i \sin{\frac{\pi (l-m)}{k}} \Bigr]
=(-1)^{\frac{M}{2}(M-1)} q^{-\frac{M}{12}(M^2 -1)} (1-q)^{\frac{M}{2}(M-1)} G_2 (M+1 ;q) ,\NN\\
&&\tanh{\frac{x_a -x_b}{2k}}
=\frac{1 -q^{\frac{i (x_b -x_a )}{2\pi}} }{1 +q^{\frac{i(x_b -x_a )}{2\pi}} } , \quad
\prod_{l=0}^{M-1} \tanh{\frac{x_j +2\pi i (l+1/2)}{2k}} 
=(-1)^M  \frac{(q^{\frac{i x_a}{2\pi} +\frac{1}{2}})_M}{(-q^{\frac{i x_a}{2\pi} +\frac{1}{2}})_M} ,\NN
\end{\eqa}
and making a transformation $s_a =i y_a -\pi$, the partition function takes the form of
\begin{\eqa}
Z_{\rm ABJ}
&=& \frac{i^{-\frac{1}{2}(N^2 +(N+M)^2 ){\rm sign}(k) +N+\frac{M}{2}}  
             (-1)^{\frac{1}{2}N(N-1)}}{2^N k^{N+M/2}N!} (1-q)^{\frac{M(M-1)}{2}} G_2 (M+1 ;q) \NN\\
&& \int_{-i \infty -2\pi\eta}^{i \infty -2\pi\eta} \frac{d^N s}{(-2\pi i )^N} 
\prod_{a=1}^N \frac{1}{2\sin{\frac{s_a}{2}}} \frac{(q^{\frac{s_a}{2\pi} +1})_M}{(-q^{\frac{s_a}{2\pi} +1})_M} 
\prod_{1\leq a<b \leq N}\frac{(1 -q^{\frac{s_b -s_a }{2\pi}} )^2}{(1 +q^{\frac{s_b -s_a }{2\pi}} )^2} .
\end{\eqa}
Although $\eta$ is naively $\eta =1/2$,
we can change $\eta$ in a range $0<\eta <1$ by the Cauchy integration theorem.
If we take $\eta =0_+$ for $2M\leq |k|$ and $\eta =\frac{1}{2}+0_+$ for $2M=|k|+1$, 
then this is nothing but the AHS formula for $2M\leq |k|$ and $2M = |k|+1$ (as a special case of $2M \geq |k|$), respectively.
Note that the choice \eqref{eta_choice} of $\eta$ is still correct for general $N$.

For a later convenience, we introduce the $U(M)_k$ pure Chern-Simons partition function (without level shift) on $S^3$ 
as \cite{Marino:2002fk,Marino:2004uf}
\begin{\eqa}
Z_{\rm CS}^{(M)}(k)
&=& i^{-{\rm sign}(k)\frac{M^2}{2} } i^{\frac{M}{2}} k^{-\frac{M}{2}} q^{-\frac{M}{12}(M^2 -1)} (1-q)^{\frac{M}{2}(M-1)} G_2 (M+1;q) \NN\\
&=&  |k|^{-\frac{M}{2}}  \prod_{l=1}^{M-1} \left( 2\sin{\frac{\pi l}{|k|} } \right)^{M-l}  .
\end{\eqa}
In terms of this, the ABJ partition function can be rewritten as
\begin{\eqa}
Z_{\rm ABJ}
&=&  \frac{i^{-{\rm sign}(k)N(N -1) } (-1)^{\frac{N}{2}(N-1)}    }{ N! 2^N |k|^N} 
 q^{\frac{M}{12}(M^2 -1)} Z_{\rm CS}^{(M)}(k) \NN\\
&& \int_{-i \infty -2\pi \eta}^{i \infty -2\pi \eta} \frac{d^N s}{(-2\pi i )^N} 
\prod_{a=1}^N \frac{1}{2\sin{\frac{s_a}{2}}}  \prod_{l=1}^M \tan{\frac{s+2l\pi }{2|k|}}
\prod_{1\leq a<b \leq N}\frac{(1 -q^{\frac{s_b -s_a }{2\pi}} )^2}{(1 +q^{\frac{s_b -s_a }{2\pi}} )^2} .
\label{rewrite}
\end{\eqa}

\subsubsection*{Remark}
We can also express the ABJ partition function as
\begin{\eqa}
Z_{\rm ABJ}^{(N,N+M)}(k) 
&=&  \frac{e^{-\frac{iM \pi}{4} {\rm sign}(k) } q^{\frac{M}{12}(M^2 -1)} \mathcal{N}_{\rm ABJ} }{N! |k|^{N+\frac{M}{2}}}  
\int_{-\infty}^\infty \frac{d^{N+M} y}{(2\pi )^{N+M}}   
   \frac{ \prod_{j<l}2\sinh{\frac{y_j -y_l }{2k}} \prod_{a<b}2\sinh{\frac{y_a -y_b }{2k}}}{\prod_{j,b}2\cosh{\frac{y_j -y_b }{2k}}} \NN\\
&& \prod_{a=1}^N \frac{1}{2\cosh{\frac{y_a}{2}}} 
  \prod_{l=N+1}^{N+M} \Biggl[ 2\pi    \delta \left( y_l -\frac{2\pi}{i}(N+M+1/2 -l)  \right) \Biggr] .
\label{brane}
\end{\eqa}
Each factor in this integrand has an interpretation from the brane picture.
Recall that the type IIB brane construction for the $U(N)_k \times U(N+M)_{-k}$ ABJ theory 
consists of $N$ circular D3-branes, $(1,-k)$5-brane, NS5-brane and $M$ D3-branes suspended between the two 5-branes \cite{Aharony:2008gk}. 
Taking S-transformation, the $(1,-k)$5-brane and NS5-brane become $(-k,1)$5-brane and D5-brane, respectively.
First, note that the second factor
\[
\prod_{a=1}^N \frac{1}{2\cosh{\frac{y_a}{2}}}
\]
agrees with the contribution from a bi-fundamental hypermultiplet \cite{Kapustin:2009kz}.
This multiplet comes from strings ending on the D5-brane and D3-branes.
Next, the first factor
\begin{\eq}
\frac{1}{N! |k|^{N+\frac{M}{2}}}
   \frac{ \prod_{1\leq j<l\leq N+M}2\sinh{\frac{y_j -y_l }{2k}} \prod_{1\leq a<b \leq N}2\sinh{\frac{y_a -y_b }{2k}}}
          {\prod_{j=1}^{N+M}\prod_{b=1}^N 2\cosh{\frac{y_j -y_b }{2k}}} 
\label{1k5}
\end{\eq}
is a bit nontrivial.
For $M=0$, the authors in \cite{Gulotta:2011si} argued that
this factor comes from a system of the $(1,-k)$5-brane and D3-branes (see also \cite{Benvenuti:2011ga}).
Hence we can interpret \eqref{1k5} as natural generalization of this contribution.
Finally, the last factor
\[
\prod_{l=N+1}^{N+M}   \delta \left( y_l -\frac{2\pi}{i}(N+M+1/2 -l)  \right) 
\]
reflects a fact that the $M$ suspended D3-branes are locked into position by the two 5-branes.

\subsection{For $2M \geq |k|-1$}
\label{sec:latter}
The integration \eqref{turning} is apparently divergent for this case.
Instead of imposing some regularizations, we use the Seiberg-like duality \eqref{Seiberg} for the ABJ theory \cite{Aharony:2008gk}.
Note that this duality has been already proven for the $S^3$ partition functions
because the duality comes \cite{Kapustin:2010mh} from the Giveon-Kutasov duality \cite{Giveon:2008zn} proven in \cite{Willett:2011gp}.
Since the dual ABJ partition function is given by \eqref{rewrite} for $2M \geq |k|-1$,
we can express $Z_{\rm ABJ}^{(N,N+M)}(k)$ in terms of $Z_{\rm ABJ}^{(N+|k|-M,N)}(-k)$  
through the duality. 

Let us show $Z_{\rm ABJ} =Z_{\rm AHS}$
for\footnote{
For $2M=|k|$ and $2M=|k|+1$, we can also apply the argument in sec.~\ref{sec:former}.
We set the condition $2M \geq |k|-1 $ such that
the dual ABJ partition function is given by \eqref{rewrite}.
} $2M \geq |k|-1$.
Via the Seiberg-like duality \eqref{Seiberg} as mathematical identity, the ABJ partition function is given by
\begin{\eqa}
Z_{\rm ABJ}^{(N,N+M)}(k) 
&=&(-1)^{\frac{N+M}{2}(N+M-1) +\frac{N+|k|-M}{2}(N+|k|-M-1)} 
    i^{-\frac{1}{2}(2N^2 -(N+M)^2 -(N+|k|-M)^2 ){\rm sign}(k)} \NN\\
&&  e^{\frac{\pi i}{12}(k^2 +6N|k| -6|k| +2 ){\rm sign}(k)} Z_{\rm ABJ}^{(N,N+|k|-M)} (-k) .
\end{\eqa}
Here the phase factor has been determined\footnote{
Note that our normalization for the partition function differs from the original paper \cite{Kapustin:2010mh} by
$Z_{\rm ours}^{(N,N+M)}(k)=(-1)^{\frac{N}{2}(N-1)+\frac{N+M}{2}(N+M-1)}\mathcal{N}_{\rm ABJ}Z_{\rm KWY}^{(N,N+M)}(k)$.
} in \cite{Kapustin:2010mh}.
Plugging \eqref{rewrite} into this leads us to
\begin{\eqa}
Z_{\rm ABJ}^{(N,N+M)}(k) 
&=&  \frac{i^{-{\rm sign}(k)N(N -1) } (-1)^{\frac{N}{2}(N-1)}    }{ N! 2^N |k|^N} 
 q^{\frac{M}{12}(M^2  -1)} Z_{\rm CS}^{(|k|-M)}(-k) \NN\\
&& \int_{-i \infty -0_+}^{i \infty - 0_+} \frac{d^N s}{(-2\pi i )^N} 
\prod_{a=1}^N \frac{1}{2\sin{\frac{s_a}{2}}}  \prod_{l=1}^{|k|-M} \tan{\frac{s+2l\pi }{2|k|}}
\prod_{1\leq a<b \leq N}\frac{(1 -q^{\frac{s_b -s_a }{2\pi}} )^2}{(1 +q^{\frac{s_b -s_a }{2\pi}} )^2} .\NN
\end{\eqa}
By using the level-rank duality\footnote{See e.g. appendix.~B of \cite{Kapustin:2010mh} for a proof.} 
for the pure CS theory: $Z_{\rm CS}^{(|k|-M)}(-k) = Z_{\rm CS}^{(M)}(k)$ 
and eq.~(E.2) in \cite{Awata:2012jb}:
\begin{\eq}
\frac{1}{\sin{\frac{s}{2}}} \prod_{l=1}^M \tan{\frac{s+2\pi l}{2|k|}}
=\left. \frac{1}{\sin{\frac{s}{2}}} \prod_{l=1}^M \tan{\frac{s+2\pi l}{2|k|}} \right|_{s\rightarrow s+2\pi \left( -\frac{|k|}{2} +M \right), M\rightarrow |k|-M} ,
\end{\eq}
we obtain
\begin{\eqa}
Z_{\rm ABJ}^{(N,N+M)}(k) 
&=& \frac{i^{-\frac{1}{2}(N^2 +(N+M)^2 ){\rm sign}(k) +N+\frac{M}{2}}  
             (-1)^{\frac{1}{2}N(N-1)}}{2^N k^{N+M/2}N!} (1-q)^{\frac{M(M-1)}{2}} G_2 (M+1 ;q) \NN\\
&& \int_{-i \infty -2\pi \eta}^{i \infty -2\pi \eta} \frac{d^N s}{(-2\pi i )^N} 
\prod_{a=1}^N \frac{1}{2\sin{\frac{s_a}{2}}} \frac{(q^{\frac{s_a}{2\pi} +1})_M}{(-q^{\frac{s_a}{2\pi} +1})_M} 
\prod_{1\leq a<b \leq N}\frac{(1 -q^{\frac{s_b -s_a }{2\pi}} )^2}{(1 +q^{\frac{s_b -s_a }{2\pi}} )^2} ,
\end{\eqa}
where
\begin{\eq}
\eta = -\frac{|k|}{2}+M +0_+ .
\end{\eq}
For $2M\geq |k|$, this exactly agrees with $Z_{\rm AHS}$.
For $2M=|k|-1$,
we can show that
making a transformation $s_a \rightarrow -s_a$ with a choice $\eta =0_-$ and using the periodicity of the integrand: $s_a \sim s_a +4k\pi i$
give the AHS formula.
Thus we find again that the choice \eqref{eta_choice} of the integral contour is valid for general $N$.

\subsubsection*{Remark}
As already discussed in \cite{Awata:2012jb}, 
the ABJ partition function vanishes for $M>k$ 
since the pure CS partition function in the prefactors vanishes for this case.
This manifests an expectation that the supersymmetries are spontaneously broken in this case \cite{Aharony:2008gk,Kitao:1998mf} (see also \cite{Hashimoto:2010bq}).

\section{Discussion}
\label{sec:con}
In this paper we have proven that
the ABJ partition function on $S^3$ is exactly the same as the formula \eqref{AHS}
recently proposed by Awata, Hirano and Shigemori \cite{Awata:2012jb},
which can be interpreted as the ``mirror'' description of the ABJ partition function.
It has also turned out that  
the choice \eqref{eta_choice} of the integral contour, previously determined only for $N=1$, 
is still correct for general $N$.
Our proof heavily relied on the determinant identity \eqref{det}
and the following illuminating structure:
\begin{\eq}
Z_{\rm ABJ}
\sim \int d^{N+M} \mu d^N \nu
\Bigl[ \langle M | b (e^{\mu_1} ) \cdots b(e^{\mu_{N+M}}) c(-e^{\nu_N} ) \cdots c(-e^{\nu_1} ) | 0\rangle \Bigr]^2
e^{-\frac{ik}{4\pi}\left( \sum_i \mu_i^2 -\sum_a \nu_a^2 \right) } ,
\end{\eq}
which might be reminiscent of the AGT relation \cite{Alday:2009aq}.
This would imply that
somehow the Boson-Fermion correspondence knows how to simplify the ABJ partition function on $S^3$.
One can see similar structure also in general $\mathcal{N}=3$ quiver CSM.
It is interesting if we find any physical origin of this structure.
A recent work \cite{Nieri:2013yra} in three dimensions and topological string perspective \cite{Aganagic:2003qj}
might provide valuable insights along this direction.

For $2M>|k|+1$, we have to change the integral contour.
We can also express the partition function for this case as 
\begin{\eqa}
\left. Z_{\rm ABJ} \right|_{2M>|k|+1} 
&\sim &
\int_{-\infty}^\infty d^{N+M} y   
   \frac{ \prod_{j<l}2\sinh{\frac{y_j -y_l }{2k}} \prod_{a<b}2\sinh{\frac{y_a -y_b }{2k}}}{\prod_{j,b}2\cosh{\frac{y_j -y_b }{2k}}}
   \prod_{a=1}^N \frac{1}{2\cosh{\frac{y_a}{2}}} \NN\\
&& \left.  
  \prod_{l=N+1}^{N+M} \Biggl[ 2\pi    \delta \left( y_l -\frac{2\pi}{i}(N+M+1/2 -l)  \right) \Biggr] 
   \right|_{y_a \rightarrow y_a +2\pi i \left( -\frac{|k|}{2} +M   \right) } .\NN
\end{\eqa}
The similar representation \eqref{brane} was useful to find the brane interpretation for $2M\leq |k|+1$.
Formally this shift is similar to anomalous R-charge, imaginary mass and voertex loop \cite{Kapustin:2009kz}. 
It would be intriguing to interpret this shift in the integrand from the brane picture.

As mentioned in sec.~\ref{sec:intro},
the AHS representation \eqref{AHS} is suitable for the Fermi gas approach and Monte Carlo simulation,
and easier to study the higher spin limit 
than the original representation \eqref{starting}.
Therefore it is illuminating to apply these approaches to the ABJ partition function and investigate wide parameter region.

Finally we comment on a relation between the ABJ theory and $L(2,1)$ matrix model.
The authors in \cite{Awata:2012jb} obtained \eqref{AHS} by
an analytic continuation from the partition function of the $L(2,1)$ matrix model represented by a convergent series.
Because its analytic continuation to the ABJ theory yields an ill-defined series,
the AHS formula corresponds to its well-defined integral representation 
and correctly reproduces the formal series order by order in the perturbative expansion.
Since here we have proven $Z_{\rm ABJ}=Z_{\rm AHS}$,
our argument combined with the one in \cite{Awata:2012jb}
might give key idea for rigorous proof of the analytic continuation.

\subsection*{Acknowledgment}
We would like to thank Hidetoshi Awata, Shinji Hirano, Keita Nii and Masaki Shigemori 
for responses to our questions about the work \cite{Awata:2012jb} and another related work in progress.
We would also like to thank Kazuo Hosomichi, Marcos Mari\~no, Sanefumi Moriyama and Kazumi Okuyama for valuable discussions and comments.
We are grateful to Yukawa Institute for Theoretical Physics
for hospitality, where part of this work was done.


\providecommand{\href}[2]{#2}\begingroup\raggedright\endgroup

\end{document}